 \documentclass[twocolumn]{aastex6}
\shorttitle{Feeding and feedback in 3C~120}
\shortauthors{Tombesi et al.}
\begin{document}


%\title{The Chandra high energy resolution X-ray view of feeding and feedback in the powerful radio galaxy 3C~120}

\title{Feeding and feedback in the powerful radio galaxy 3C~120}

%% Use \author, \affil, plus the \and command to format author and affiliation 
%% information.  If done correctly the peer review system will be able to
%% automatically put the author and affiliation information from the manuscript
%% and save the corresponding author the trouble of entering it by hand.
%%
%% The \affil should be used to document primary affiliations and the
%% \altaffil should be used for secondary affiliations, titles, or email.
%% Authors with the same affiliation can be grouped in a single

\author{F. Tombesi$^{1,2,3}$,  R.~F. Mushotzky$^2$, C.~S. Reynolds$^2$,
  T. Kallman$^{1}$, J.~N. Reeves$^{4}$, V. Braito$^{5}$,
  Y. Ueda$^{6}$, M.~A. Leutenegger$^1$, B.~J. Williams$^1$, {\L}. Stawarz$^7$ and M. Cappi$^8$}
\affil{$^{1}$X-ray Astrophysics Laboratory, NASA/Goddard Space Flight Center, Greenbelt, MD 20771, USA; francesco.tombesi@nasa.gov}
\affil{$^{2}$Department of Astronomy, University of Maryland, College Park, MD 20742, USA; ftombesi@astro.umd.edu}
\affil{$^3$Dipartimento di Fisica, Universit\`{a} di Roma Tor Vergata, Via della Ricerca Scientifica 1, I-00133 Roma, Italy}
\affil{$^4$Astrophysics Group, School of Physical and Geographical Sciences, Keele University, Keele, Staffordshire, ST5 5BG, UK}
\affil{$^5$INAF - Osservatorio Astronomico di Brera, Via Bianchi 46 I-23807 Merate (LC), Italy}
\affil{$^6$Department of Astronomy, Kyoto University, Kyoto 606-8502, Japan}
\affil{$^7$Astronomical Observatory, Jagiellonian University, ul. Orla 171, 30-244, Krak\'ow, Poland}
\affil{$^8$INAF-IASF Bologna, Via Gobetti 101, I-40129 Bologna, Italy}


%\altaffiltext{1}{AAS Journals Data Scientist}
%\altaffiltext{2}{greg.schwarz@aas.org}
%\altaffiltext{3}{AAS Journals Associate Editor-in-Chief}
%\altaffiltext{4}{AAS Director of Publishing}
%\altaffiltext{5}{IOP Senior Publisher for the AAS Journals}

\begin{abstract}

%The first lesson in the tutorial is to remind authors that the AAS
%Journals, the Astrophysical Journal (ApJ), the Astrophysical Journal
%Letters (ApJL), and Astronomical Journal (AJ), all have a 250 word limit
%for the abstract.  If you exceed this length the Editorial office will ask
%you to shorten it.

%{****\bf CONSIDER ANOTHER PAPER FOR 3C111??? LETTER?? INCLUDE MOLECULAR
%  CLOUD DISCUSSION? NEED UPDATE TABULATED VALUE NH!!!!****}

%{\bf HIGH TEMPERATURE KEV FRIED GAS, NO WAR ABSORBER TOO HOT!!! 3C120}

%{\bf BROAD FEATURE, VELOCITY, TURBULENCE!!!???}

%{\bf PUBLISH ZEROTH IMAGE AND RADIO NOW!? ASK PERLMAN!?}

%{\bf SEND PAPER VEILLEUX, FERUGLIO ASK ALMA OR IRAM OBSERVATION
%  3C120?? MOLECULAR GAS Mazzarella et al. 1993, OTHER PAPERS???}

%{\bf DISCUSS MOLECULAR GAS OUTFLOW! MERGER! DISCUSSION, INTRODUCTION,
%  ABSTRACT AND CONCLUSION!!!<<<----}

%A RECOLLIMATION SHOCK 80 MAS FROM THE CORE IN THE JET OF THE RADIO GALAXY 3C 120: OBSERVATIONAL EVIDENCE AND MODELING

We present the spectral analysis of a 200~ks observation of the
broad-line radio galaxy 3C~120 performed with the high energy
transmission grating (HETG) spectrometer on board the \emph{Chandra} X-ray Observatory.
We find (i) a neutral absorption component
intrinsic to the source with column density of $\text{log}N_H = 20.67\pm0.05$~cm$^{-2}$, (ii) no evidence for a warm absorber with
an upper limit on the column density of just $\text{log}N_H < 19.7$~cm$^{-2}$ assuming the typical ionization parameter log$\xi$$\simeq$2.5~erg~s$^{-1}$~cm, the warm
absorber may instead be replaced by (iii) a hot emitting gas with temperature $kT
\simeq 0.7$~keV observed as soft X-ray emission from ionized Fe
L-shell lines which may originate from a kpc scale shocked bubble
inflated by the AGN wind or jet with a shock velocity of about 1,000~km~s$^{-1}$ determined by the emission line width, (iv) a neutral Fe
K$\alpha$ line and accompanying emission lines indicative of a
Compton-thick cold reflector with low reflection fraction
$R\simeq0.2$, suggesting a large opening angle of the torus, (v) a
highly ionized Fe~XXV emission feature indicative of photoionized gas with ionization parameter
log$\xi$$=$$3.75^{+0.27}_{-0.38}$~erg~s$^{-1}$~cm 
and a column density of $\text{log}N_H > 22$~cm$^{-2}$ localized
within $\sim$2~pc from the X-ray source, and (vi) possible signatures
for a highly ionized disk wind. Together with previous evidence for intense
molecular line emission, these results indicate that 3C~120 is likely a
late state merger undergoing strong AGN feedback.

%These results and the facts that the radio galaxy 3C~120 has
%borderline radio classification between FRI and FRII showing a one-sided low
%inclination relativistic jet, its host galaxy is a
%disturbed S0 galaxy, and it has been detected in the
%$\gamma$-rays, make this source a very interesting bridge between blazars,
%radio galaxies, and Seyferts.

%{\bf ****ZERO-TH ORDER***}
%{\bf BLOB IN 3C 120!!!???? CHECK!!!}
%{\bf X-RAY JET CHANDRA WEBSITE!!}
%{\bf PREVIOUS OBSERVATION 3C 111 CHANDRA ACIS/HETG IN 2004! Check
% variability! Harris et al. 2004}
%{\bf COMBINE WITH RADIO, HUBBLE IMAGE? ERIC PERLMAN, CLAUTICE!!
%  ***Siemiginowska Aneta *** <====}
%{\bf Massaro et al. 2011, paper radio jets sample? }
%NEW PROPOSAL, CHECK RADIO IMAGE FOR JET!!! ZEROTH ORDER
%{\bf 3C120 backward shock redshifted 3000km/s?? Change jet direction about 5arcsec! Optocal interaction arm Galaxy???}
%3C120 MOLECULAR OUTFLOW JET INTERACTION???
%{\bf COLD GAS INFLOW IN 3C120!!?? S0 Galaxy!!! OPTICAL OBS? 0.1keV... CHECK MCNAMARA!!! TALK GSFC}

%{\bf ****CHEC PAPER, YAQOOB; OGLE ET AL 2005, TORRESI ET AL. 2012; MCKERNAN ET AL. LINE IN XMM, RGS DATA!!!??? Analize Again???****}

%{\bf 3C120 IS A SPIRAL, SHOULD NOT HAVE HOT GAS -> IN NED CLASSIFIED AS “S0”, this is intermediate between Elliptical and Spiral!!! EVEN MORE INTERESTING!!!}

%{\bf ****CONNECTION BETWEEN INCLINATION AND DETECTION WARM ABSORBERS??? FAR OUT, DETECTABLE BETTER FOR HIGH INCLINATION, HIGHER COLUMN!!!! 3C120 LOWEST INCLINATION ANGLE!!! 3C445 HIGHES!!! ESTIMATE THETA DEPENDENCE column density, velocity, ionization, compare with FUKUMURA!!!<---}

\end{abstract}


\section{Introduction} 

Increasing evidence points toward the possibility that virtually every galaxy hosts a supermassive black hole
(SMBH) at its core, and that the evolution of both the SMBH and the
galaxy's baryonic component, i.e. stars and the intergalactic medium, may be intimately linked. The origin of this co-evolution
is highly debated but it is likely that active galactic nuclei (AGN)
may play an important role through a process known as ``feedback'' (e.g., Silk \& Rees
1998; Fabian 1999; Springel et al.~2005; Croton et
  al.~2006; Hopkins et al.~2008; Tombesi et al.~2015). Depending on whether the source is
radio-quiet or radio-loud, the feedback is usually referred to as
quasar/wind mode or radio/jet mode, respectively (e.g., Fabian 2012; King \& Pounds 2015). 

The dichotomy between radio-quiet and radio-loud AGN is still not
fully understood, but it seems that the latter are more preferentially found in elliptical/disturbed
galaxies and minor mergers (e.g., Xu, Livio \&
Baum 1999; Sikora, Stawarz \& Lasota 2007; Chiaberge et al.~2015). The black hole spin may
also play a role in driving the most powerful jets (e.g., Wilson \&
Colbert 1995; Hughes \& Blandford 2003; Sikora, Stawarz \& Lasota
2007; Garofalo, Evans \& Sambruna 2010; Tchekhovskoy, Narayan \&
McKinney 2010). Moreover, the accumulation of
magnetic flux near the black hole horizon has been suggested to play a
role in the radio-loud/radio-quiet dichotomy as well (e.g., Tchekhovskoy, Narayan \&
McKinney 2011; Sikora \& Begelman 2013). On the other hand, deeper observations, especially in the X-ray band, are showing
that the differences in the central engines between the two classes of
sources are rather subtle (Hardcastle et al.~2009; Tazaki et al.~2013; Bostrom et
al. 2014). 

Sensitive X-ray observations of radio galaxies show evidence for winds in this class of sources, similar to
those observed in Seyferts, which are radio-quiet as a class (Tombesi et al.~2010a, 2013a; Gofford et
al.~2013). Warm absorbers (WA) have been reported in 3C~382, 3C~445,
3C~390.3 and 4C$+$74.26 (Ballantyne 2005; Reeves et al.~2009, 2010;
Torresi et al. 2010, 2012; Tombesi et al.~2016). Moreover, more
extreme ultrafast outflows (UFOs) have also been reported in 3C~111,
3C~120, 3C~390.3, 3C~445, 4C$+$74.26, and Cygnus A (Tombesi et
al.~2010b, 2011; Ballo et al.~2011; Braito et al.~2011; Gofford et
al.~2013; Reynolds et al.~2015). A recent X-ray study of a sample of
26 radio galaxies reported that the frequency of UFOs is likely in the
interval $f$$\simeq (50\pm20)$\% (Tombesi et al.~2014). Thus, contrary
to the jet dichotomy, it seems that some type of winds may be present
in both radio-quiet and radio-loud AGNs. Some of these winds may be
powerful enough to provide a concurrent contribution to AGN feedback
with jets (Tombesi et al.~2012a, b; Gofford et al.~2015). 

Spectral features indicating reflection from the inner
accretion disk and the parsec scale torus or broad line region have
been reported in several radio galaxies as well (e.g., Kataoka et al. 2007; Sambruna et
al.~2009; Tombesi et al.~2013b; Tazaki et al.~2013; Bostrom et
al. 2014; Lohfink et al.~2015). Indeed, the presence of
  signatures of the accretion disk, winds, and jets in the composite broad-band spectra of broad-line radio galaxies (BLRGs) make them the ideal
objects to study the interplay among these components (e.g., Marscher
et al. 2002; Chatterjee et al. 2009, 2011; Tombesi et al.~2011, 2012,
2013b; Lohfink et al.~2013; Fukumura et al.~2014; Clautice et al.~2016).

Here, we show the analysis of a long 200~ks \emph{Chandra} high energy
transmission grating (HETG) spectrometer observation of the radio
galaxy 3C~120 ($z = 0.033$). This is the second paper on this series, the previous one was focused on the radio
galaxy 3C~390.3 (Tombesi et al.~2016). The combined high energy
resolution and moderate sensitivity in the wide energy band
E$=$0.5--7~keV provided by the \emph{Chandra} HETG spectrum are crucial for the
detection of possible emission and absorption features from a wide
range of ionization species, and it will provide one of the best high
energy resolution spectrum of a radio galaxy to date.

The radio galaxy 3C~120 is particularly interesting for several
different reasons: it has a
borderline classification between FRI and FRII exhibiting a one-sided
low inclination jet with bright X-ray knots (Harris et al.~2004), superluminal knots have been
detected at parsec scales in the radio jet (G{\'o}mez et al.~2000), it has been detected in the
$\gamma$-rays (Kataoka et al.~2011), it showed puzzling emission
features in the soft X-rays (Petre et al.~1984; Torresi et al.~2012;
Tombesi et al.~2014), its host galaxy is a disturbed S0 galaxy
possibly the remnant of a minor merger, and it also shows significant
molecular gas emission (Mazzarella et al.~1993;
Evans et al.~2005). Therefore, this source may well represent a bridge between blazars,
radio galaxies, Seyferts, and the final merger state of ultraluminous
infrared galaxies (ULIRGs).

\floattable 
%\begin{deluxetable}{lcccccccc}[t!]
\begin{deluxetable}{lcccc}[ht!]
\tablecaption{Chandra HETG observations log for 3C~120.}
\tablehead{
\colhead{Obs} & \colhead{ID} & \colhead{Date} & \colhead{Exp} & \colhead{Rate}
}
\startdata
%\colnumbers
%\hline
%\multicolumn{5}{c}{3C~120}\\
%\hline
1 & 16221 & 2014/12/19 & 78 & 0.82/0.40\\
2 & 17564 & 2014/12/22 & 30 & 0.77/0.38\\
3 & 17565 & 2014/12/27 & 43 & 0.53/0.26\\
4 & 17576 & 2015/01/27 & 43 & 0.82/0.41\\
%\hline
%\multicolumn{5}{c}{3C~111}\\
%\hline
%1 & 16219 & 2014/11/04 & 143 & 0.47/0.28\\
\enddata
%\tablenotetext{a}{Adjusted for inflation}
%\tablenotetext{b}{Accounts for the change from page charges to digital quanta in April, 2011}
\tablecomments{Columns: observation number; observation ID; observation date; exposure in ks; MEG/HEG count rates.}
\end{deluxetable}

\section{Data reduction and analysis}

Here, we describe the analysis of the \emph{Chandra} HETG spectrum of the
radio galaxy 3C~120. The observation is composed of four exposures
performed within one month between December 2014 and January 2015 for
a total exposure of about 200~ks, see
Table~1 for details.  No significant spectral
variability is observed between the four exposures and the spectra are
consistent with only $\simeq$3\% and $\simeq$10\% variations in photon
index and source count rate, respectively. 

The spectra were extracted using the \emph{CIAO}
package v4.7 and the associated CALDB. Only the first order dispersed
spectra were considered for both the Medium Energy Grating (MEG) and
High Energy Grating (HEG), and the $\pm$1 orders for each grating were
subsequently combined for each sequence. The spectra from
the separated exposures were combined to yield a single 1st order spectrum
for each of the MEG and HEG gratings. The background count-rate is
found to be always negligible. The resultant spectra were binned to the full
width half maximum (FWHM) of their spectral resolution, which 
corresponds to $\Delta\lambda = 0.023$ \AA\, and $\Delta\lambda =
0.012$ \AA\, bins for MEG and HEG, respectively. The MEG and HEG
spectra were analyzed in the energy bands E$=$0.5--7~keV and
E$=$1--7.5~keV, respectively. The analysis of the background
subtracted source spectra was performed using \emph{XSPEC} v.12.8.2
and the C-statistic was applied. We performed simultaneous fits of the
MEG and HEG spectra allowing a free cross-normalization
constant, whose fitted value was always consistent with unity. All parameters are given in
the source rest frame and the errors are at the 1$\sigma$ level if not
otherwise stated. Standard Solar abundances are assumed (Asplund et al.~2009).

\subsection{Phenomenological spectral analysis of 3C~120}

We started the spectral analysis including a continuum power-law with
$\Gamma$$\simeq$1.7 ($C/\nu = 2407/2220$). A Galactic absorption of
$N_H = 1.1\times 10^{21}$~cm$^{-2}$ modeled with \emph{wabs} in
\emph{XSPEC} is included in all the fits
(Kalberla et al.~2005). Equivalent results are obtained using the
\emph{TBabs} model in \emph{XSPEC}.  An inspection of the data in Fig.~1 show the
possible presence of absorption residuals in the soft X-rays and
possible emission lines. Indeed, an additional neutral absorber
intrinsic to the source with column density of $N_H=(4.3\pm0.5)\times
10^{20}$~cm$^{-2}$ is highly required ($\Delta C/\Delta\nu =
81/1$). Equivalent results are obtained using either \emph{zwabs} or
\emph{zTBabs} models in \emph{XSPEC}. 

\subsubsection{Emission lines in the soft X-ray band}

In Fig.~2 we can see a series of possible emission features in the
energy range between E$=$0.65--0.85~keV. When modeled with Gaussian
emission lines, the rest-frame energies of the two most significant features are
at E$=$$719\pm3$~eV and E$=$$777^{+2}_{-7}$~eV, respectively. The
parameters of the lines are reported in Table~2. We consider only the
lines with a fit improvement of $\Delta C \ge 4$ for two additional
degrees of freedom, corresponding to a detection significance higher
than 90\%.

The E$\simeq$719~eV feature is relatively broad, with a width of
$\sigma$$=$$8\pm2$~eV. The closest lines\footnote{The line
  identifications are derived from the National Institute of Standards
  and Technology (NIST) database or Verner et al. (1996).} to the
first emission feature is the ionized Fe L-shell doublet of Fe~XVII, with
transitions respectively at the
energies of E$=$727eV and E$=$739~eV. Another spectral
feature close to the observed energy is O~VII radiative recombination
continuum (RRC) at E$=$739~eV
(e.g., Liedahl \& Paerels 1996). The only other
features around this energy are the L$\alpha$ and L$\beta$
fluorescence lines from neutral or lowly ionized Fe at E$=$705~eV
and E$=$719~eV, respectively. The possible alternative interpretations
of this feature are discussed in the following section using more
physical models. 

The second emission line at the energy of E$=$$777$~eV may be
associated with the Fe~XVIII 2p$\rightarrow$3s doublet at E$=$777~eV
and E$=$779~eV, or with O~VIII Ly$\beta$ at E$=$775~eV. The latter
possibility is supported by the detection of O~VIII Ly$\alpha$ at
E$\simeq$650~eV in the \emph{XMM-Newton} reflection grating
spectrometer (RGS) spectrum (Ogle et al.~2005; Torresi et
al.~2012). The \emph{XMM-Newton} RGS provided better signal-to-noise data at energies
below E$\simeq$700~eV, while \emph{Chandra} HETG is superior at higher energies.

  \begin{figure}[t!]
  \centering
   \includegraphics[width=8.5cm,height=7cm,angle=0]{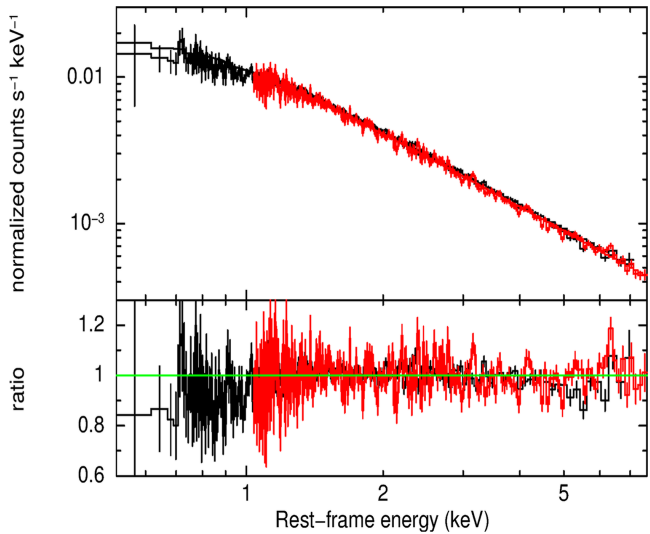}
   \caption{Combined \emph{Chandra} MEG (black) and HEG (red) spectra
     of 3C~120 compared to a Galactic absorbed power-law continuum
     model. \emph{Upper panel:} spectra and continuum model, the data
     are divided by the response effective area for each
     channel. \emph{Lower panel:} data to model ratio. We note the
     presence of soft X­ray absorption residuals and several emission
     lines. The data are binned to 4$\times$ the FWHM resolution and to a minimum signal-to-noise of 5 for clarity.}
    \end{figure}

\subsubsection{Emission lines in the Fe K band}

In Fig.~3 we show the ratio between the data and the absorbed
power-law continuum in the Fe K region between E$=$5.5--7.5~keV. We
note the presence several emission features.

The emission line at the energy of E$=$$6394\pm5$~eV is detected with very high
significance ($\Delta C/\Delta\nu = 40/2$), and it is clearly
identified with the Fe K$\alpha$ fluorescence emission doublet from
neutral or lowly ionized material (K$\alpha_1$ at E$=$6391~eV and
K$\alpha_2$ at E$=$6404~eV). The emission line at
E$=$$6232^{+8}_{-23}$~eV, even though less significant, is at the
exact energy expected for the Compton shoulder of the main Fe
K$\alpha$ line  of E$\simeq$6240~eV  (e.g., Matt 2002; Watanabe et
al.~2003; Yaqoob \& Murphy 2011). 

The emission line at E$=$$7055^{+21}_{-15}$~eV is instead at the
energy consistent with the accompanying Fe K$\beta$ fluorescence line
at E$=$7058~eV. If this interpretation is correct, then the equivalent
width of this feature is estimated to be $\simeq$60\% of the
K$\alpha$. This is much larger than the typical value of $\simeq$11\%
expected for the K$\beta$ from atomic physics (e.g., George \& Fabian
1991). One possible explanation for this discrepancy could be that the
K$\beta$ line is blended with other features. The closest emission
feature would be Fe~XXVI Ly$\alpha$ at energy of E$=$6970~eV. However,
this possibility seems unlikely because the K$\beta$ line is unresolved.

The Fe K$\beta$/Fe K$\alpha$ ratio of emission line photons that
escape from the obscuring torus may be higher than the tabulated value
when the medium is optically thick to absorption of either of the two
emission lines. This is due to the differential absorption opacities for the Fe K$\alpha$ and Fe K$\beta$ lines. Moreover, possible clumpiness in the torus may affect this ratio. In the next section we explore more physically motivated models for these features and discuss a possible interpretation for the discrepancy in the equivalent width ratio. 

The emission feature at E$=$$6703^{+5}_{-20}$~eV is consistent with
the energy of the Fe~XXV He$\alpha$ resonance emission line at
E$=$6700~eV. A similar feature was observed also in lower energy
resolution spectra from \emph{Suzaku} and \emph{XMM-Newton} (e.g.,
Tombesi et al.~2010b, 2014; Lohfink et al.~2013; Gofford et al.~2013). 
The limited signal-to-noise and bandwidth in the Fe K region of the
present \emph{Chandra} HETG spectrum does not allow to detect the
broad Fe K emission feature previously
reported in a \emph{Suzaku} observation (Kataoka et
al.~2007). The parameters of the lines are reported in Table~2.

\subsection{Spectral analysis of 3C~120 with physical models}

After the initial phenomenological modeling, we performed a more
physically motivated fit of the spectrum. The starting model was an
absorbed power-law. 

\subsubsection{Emission lines in the Fe K band}

We replaced the Fe K$\alpha$ emission line in turn with a cold
reflection component \emph{pexmon} (Nandra et al.~2007), and an
ionized reflection component \emph{xillver} (Garcia et al.~2014). We
assumed an inclination angle of 20$^{\circ}$ consistent with the radio
jet (Jorstad et al.~2005), a standard Solar abundance for iron and a high energy cut-off of
E$=$300~keV as estimated from hard X-ray observations (e.g., Kataoka
et al.~2007; Tombesi et al.~2014). The \emph{pexmon} reflection
component also includes the Fe K$\alpha$ and K$\beta$ emission lines at E$=$6.4~keV and E$=$7.058~keV with equivalent width ratio of 11\%, and the Ni K$\alpha$ at E$=$7.470~keV with a flux of $\simeq$5\% of the Fe K$\alpha$. 

The Compton shoulder of the Fe K$\alpha$ line, denoted as Fe K$\alpha_c$, is included in
\emph{pexmon}, but it is only roughly approximated as a Gaussian
emission line at the energy E$=$6.315~keV, with width $\sigma$$=$35~eV, and
equivalent width (EW) following the prescription of Matt (2002). This may be a reasonable
approximation for low resolution spectra, in which the line and
Compton shoulder are blended together, but it may be not adequate for the high resolution \emph{Chandra} HETG data discussed here. 

An initial fit comparing \emph{pexmon} to \emph{xillver} shows that
the former component is preferred, giving a reflection fraction of $R =
0.22\pm0.04$. The reflection fraction is defined as $R = \Omega/2\pi$, where $\Omega$ is the solid angle of the reflector. This is consistent with previous broad-band analyses
(e.g., Kataoka et al.~2007; Tombesi et al.~2014). The statistical
improvement of the inclusion of the \emph{pexmon} component with respect to the absorbed power-law continuum is very
high, $\Delta C/\Delta\nu = 47/1$. There is no requirement for
broadening of this component, consistent with the fact that the Fe
K$\alpha$ line is unresolved ($\sigma_E$$<$20~eV) and likely due to
material relatively distant from the SMBH. The alternative modeling with a \emph{xillver} component provides a lower statistical improvement
($\Delta C/\Delta\nu = 38/2$), with an upper limit on the ionization
parameter of log$\xi$$<$0.8 erg~s$^{-1}$~cm, again consistent with lowly
ionized or neutral material. 

In order to derive a proper modeling of the Fe K$\alpha$ emission
line, the associated Fe K$\beta$, and Compton shoulder, we replaced
\emph{pexmon} using the more detailed torus reflection model
\emph{MYTorus} (Murphy \& Yaqoob 2009; Yaqoob \& Murphy 2011). The main
parameters of this model are the power-law slope, assumed to be the
same as the power-law continuum, the inclination angle, assumed to be
$\theta$$=$20$^{\circ}$ from the radio jet (Jorstad et al.~2005), and the torus column
density. The inclusion of this component provides a very good fit,
with a statistical improvement of $\Delta C/\Delta\nu = 45/1$. The
column density is estimated to be $N_H$$>$$6\times 10^{24}$~cm$^{-2}$
at the 90\% level. Even if this fit is statistically comparable to the
one using \emph{pexmon}, in the following we will use \emph{MYTorus}
because it provides a more physically self-consistent treatment of the
X-ray reflection from cold distant material in a toroidal configuration. 

%The shape of the Compton shoulder of a fluorescent emission-line
%escaping from the torus has a dependence on the column density and
%inclination angle. It is also affected by the shape of the incident
%continuum, the covering fraction (or opening angle), and element
%abundance. The zeroth-order component of the Fe K$\alpha$ fluorescent
%emission line refers to line photons that escape without any
%interaction with the medium that they were created in. Although they
%are emitted by atoms isotropically, the angular distribution of the
%emerging zeroth-order photons will depend on the geometry, unless the
%medium is optically-thin. The zeroth-order photons constitute the
%majority of photons of the Fe K$\alpha$ emission line. The line
%photons that escape the medium after at least one interaction give
%rise to the Compton scattered component. The final scattered line profile depends on the geometry, orientation, and column density because the escape probabilities may be highly directional (Murphy \& Yaqoob 2009; Yaqoob \& Murphy 2010).

  \begin{figure}
  \centering
   \includegraphics[width=7.5cm,height=5.7cm,angle=0]{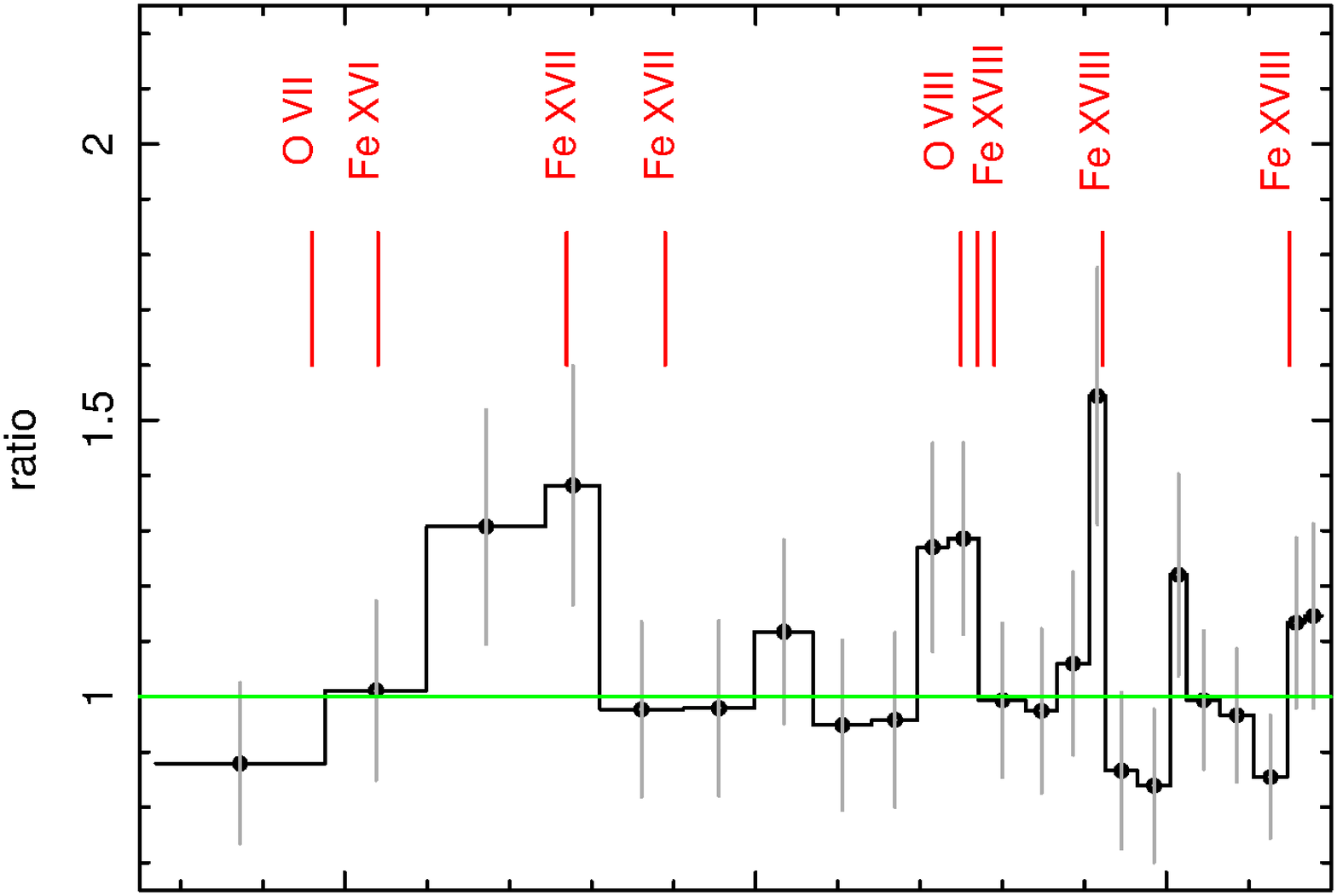}
   \includegraphics[width=7.5cm,height=5.7cm,angle=0]{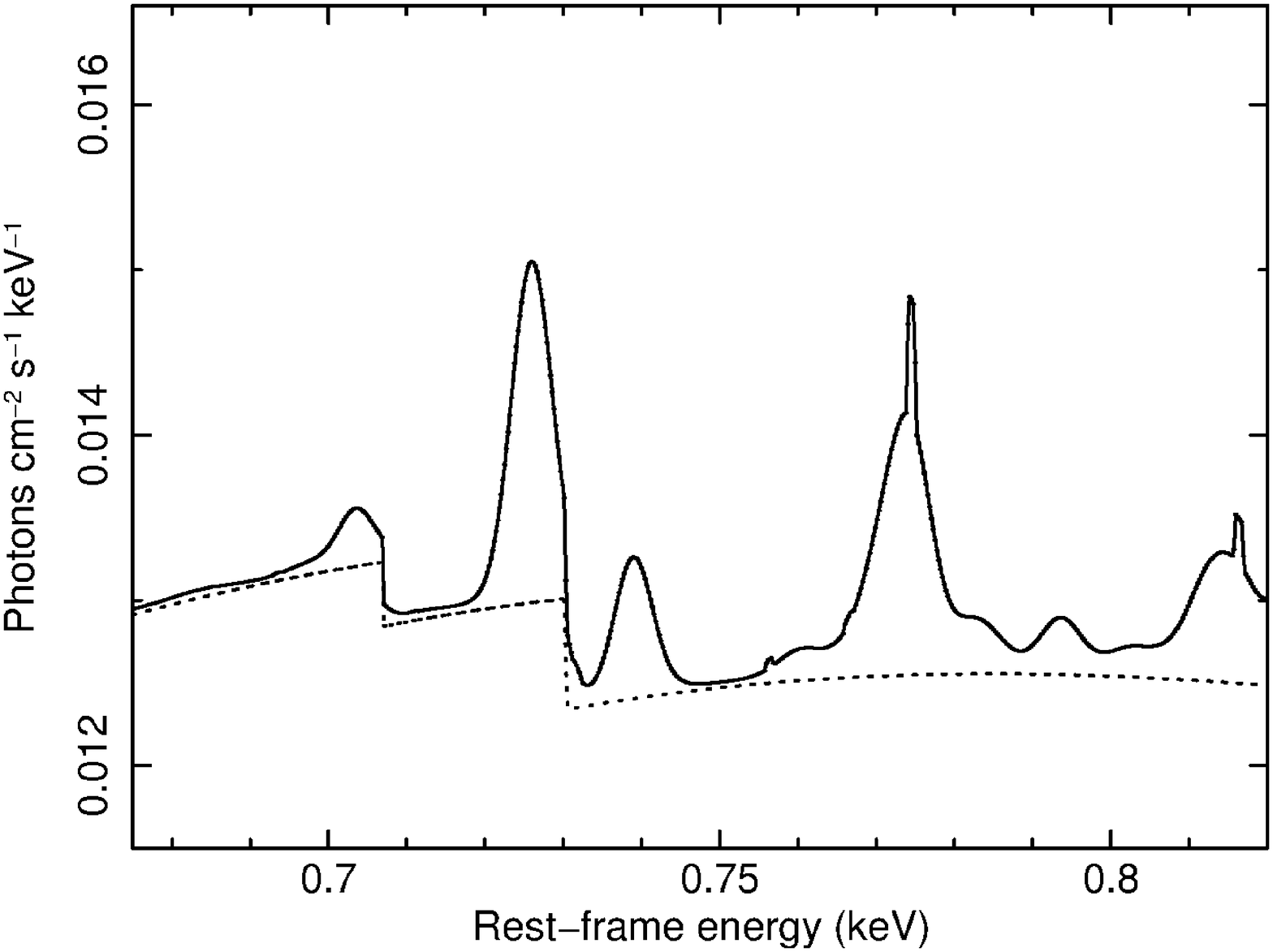}
   \caption{\emph{Upper panel:} data to model ratio of the \emph{Chandra} MEG spectrum of
     3C~120 in the E$=$0.65--0.85~keV energy band with respect to a
     Galactic absorbed power-law continuum model. The data are binned
     to 2$\times$ the FWHM resolution and a minimum signal-to-noise of
     5 for clarity. The vertical lines indicate the possible
     identifications. \emph{Lower panel:} best-fit model
     including a hot emission component for the soft X-ray lines.
    The dotted line indicates the absorbed continuum.}
    \end{figure}

The ratio between the Fe K$\alpha$ and the Compton shoulder $C = (0.1
+ 0.1 \mathrm{cos}\theta)$ depends on the inclination of the system
$\theta$ (e.g., Matt 2002). Considering an inclination consistent with
the one of the radio jet of $\theta = 20.5^{\circ}\pm 1.8^{\circ}$
(Jorstad et al.~2005) and the EW$\simeq$32~eV of the Fe K$\alpha$ line
from Table~2, we estimate an EW of the Compton shoulder of
EW$\simeq$7~eV which is consistent with the observed value. Given the
low inclination of 3C~120, the EW$\simeq$30~eV of the Fe K$\alpha$
suggests a high column density of $N_H$$>$$1\times 10^{24}$~cm$^{-2}$
for the torus, which is outside of our line of sight. This is
supported by the detection of the Compton shoulder, which indicates a
toroidal structure with a column
density of the order of $N_H$$>$$6\times 10^{24}$~cm$^{-2}$ from the
best-fit using \emph{MYTorus}. 

Considering the torus reflection model developed by Ikeda et al.~(2009) and the
calculations reported in Fig.~7 of Tazaki et al.~(2013) assuming
the parameters estimated for 3C~120, i.e., an Fe K$\alpha$ line with
EW$\simeq$30~eV, an inclination angle of $\simeq$20$^{\circ}$, and a torus
column density of $N_H$$\simeq$$1\times10^{24}$~cm$^{-2}$, we derive
that the torus should have a large opening angle of $>$80$^{\circ}$
and therefore a small covering fraction. The main difference between
the torus model by Ikeda et al.~(2009) and \emph{MYTorus} is
geometrical, with the former assuming an almost spherical geometry
with cone­shaped bipolar holes, whose opening angle is a free
parameter, and a fixed torus-like geometry for the latter. We note
that we do not observe absorption from the torus even if it is likely Compton-thick due to
the large opening angle of the torus and low inclination angle of the
disk/jet along our line of sight. 

The ionized emission line at E$=$6.7~keV can be well modeled including an
\emph{xstar} photo-ionized emission component (Kallman \& Bautista
2001). We calculated an \emph{xstar} emission table considering a
power-law continuum of $\Gamma = 1.7$. The free parameters of this model are the ionization
parameter, the column density, and the normalization. The
normalization is defined as $N$$=$$f L_{ion}/D^2$, where
$f$$=$$\Omega/4\pi$ is the covering fraction of the material,
$L_{ion}$ is the ionizing luminosity in units of $10^{38}$
erg~s$^{-1}$ from 1 (1 Ryd $=$ 13.6 eV) to 1,000 Ryd, and $D$ is the distance of the
observer to the source in kpc. The emitter normalization and column
density are degenerate. By assuming a covering fraction $f$$=$1 we can
derive a lower limit for the column density. Substituting the
appropriate values of distance\footnote{Assuming cosmological
  parameters $H_0 =  73.00$~km~sec$^{-1}$~Mpc$^{-2}$,
  $\Omega_{\text{matter}} =   0.27$, $\Omega_{\text{vacuum}} =   0.73$} $D$$\simeq$$1.35\times10^{5}$~kpc for
3C~120 and the absorption corrected ionizing luminosity of
$L_{ion}$$\simeq$$3.8\times 10^{6}(\times10^{38})$ erg~s$^{-1}$
extrapolated from the best-fit model, we obtain an estimate for the
normalization of $N$$\simeq$$2\times 10^{-4}$. The inclusion of this
component is required with a significance of 99\% ($\Delta C/\Delta\nu
= 9/2$). The resultant ionization parameter is high,
log$\xi$$=$$3.75^{+0.27}_{-0.38}$ erg~s$^{-1}$~cm. The ionization
parameter is defined as $\xi$$=$$L_\mathrm{ion}/n r^2$ erg~s$^{-1}$~cm
(Tarter, Tucker \& Salpeter 1969) where $n$ is the number density of the material, and $r$ is the distance of
the gas from the central source. We can also estimate a lower limit for the column density of $N_H$$>$$1\times 10^{22}$~cm$^{-2}$.
Even after including the cold reflection and photo-ionized emission
components we find that there is still a marginally significant
excess of emission at the energy of the neutral Fe K$\beta$
fluorescence line. The best-fit model is shown in the lower panel of Fig.~3.

%\begin{math}
%L_W = \frac{1}{2} \dot{M}_{out} v^2_{out} \sim \Omega N_H R_{in}
%v^3_{out}
%\end{math}

\subsubsection{Emission lines in the soft X-rays}

We tested several models to explain the soft X-ray emission lines. We
started by including a collisional ionization component using the
\emph{apec} model (Smith et al.~2001). We also included a Gaussian broadening given that the line
was resolved. In this case the line would be interpreted as emission
from Fe~XVII Ly$\alpha$ and slightly lower ionization species. The best-fit parameters are a broadening of
$\sigma$$=$$2.7^{+2.6}_{-0.1}$~eV, a temperature of
$kT$$=$$0.4^{+0.3}_{-0.1}$~keV and a normalization of
$N$$=$$(2.1^{+1.5}_{-1.0})\times 10^{-4}$. The fit
improvement is $\Delta C/\Delta\nu$$=$$10/3$.

  \begin{figure}
  \centering
   \includegraphics[width=7.5cm,height=5.7cm,angle=0]{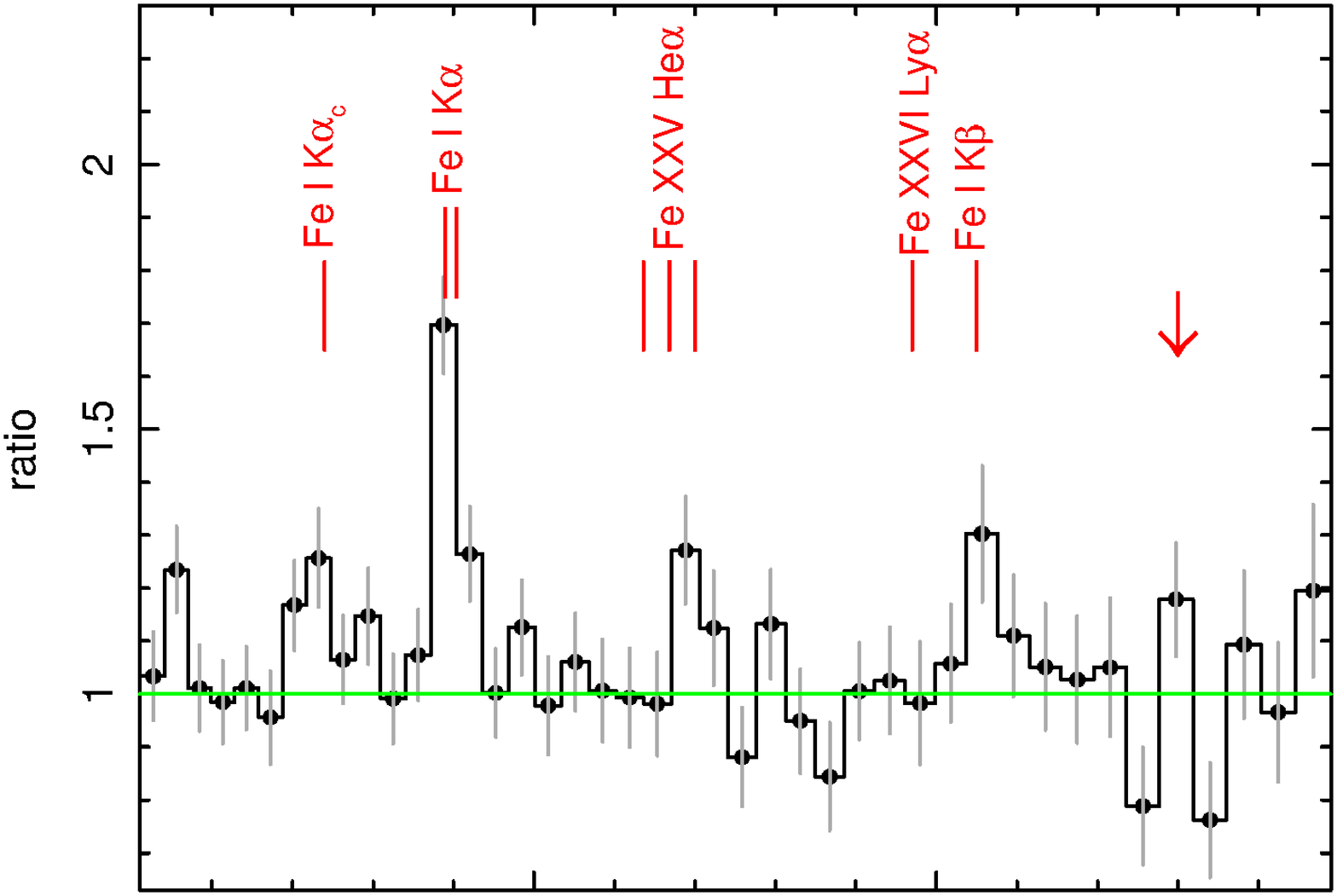}
   \includegraphics[width=7.5cm,height=5.7cm,angle=0]{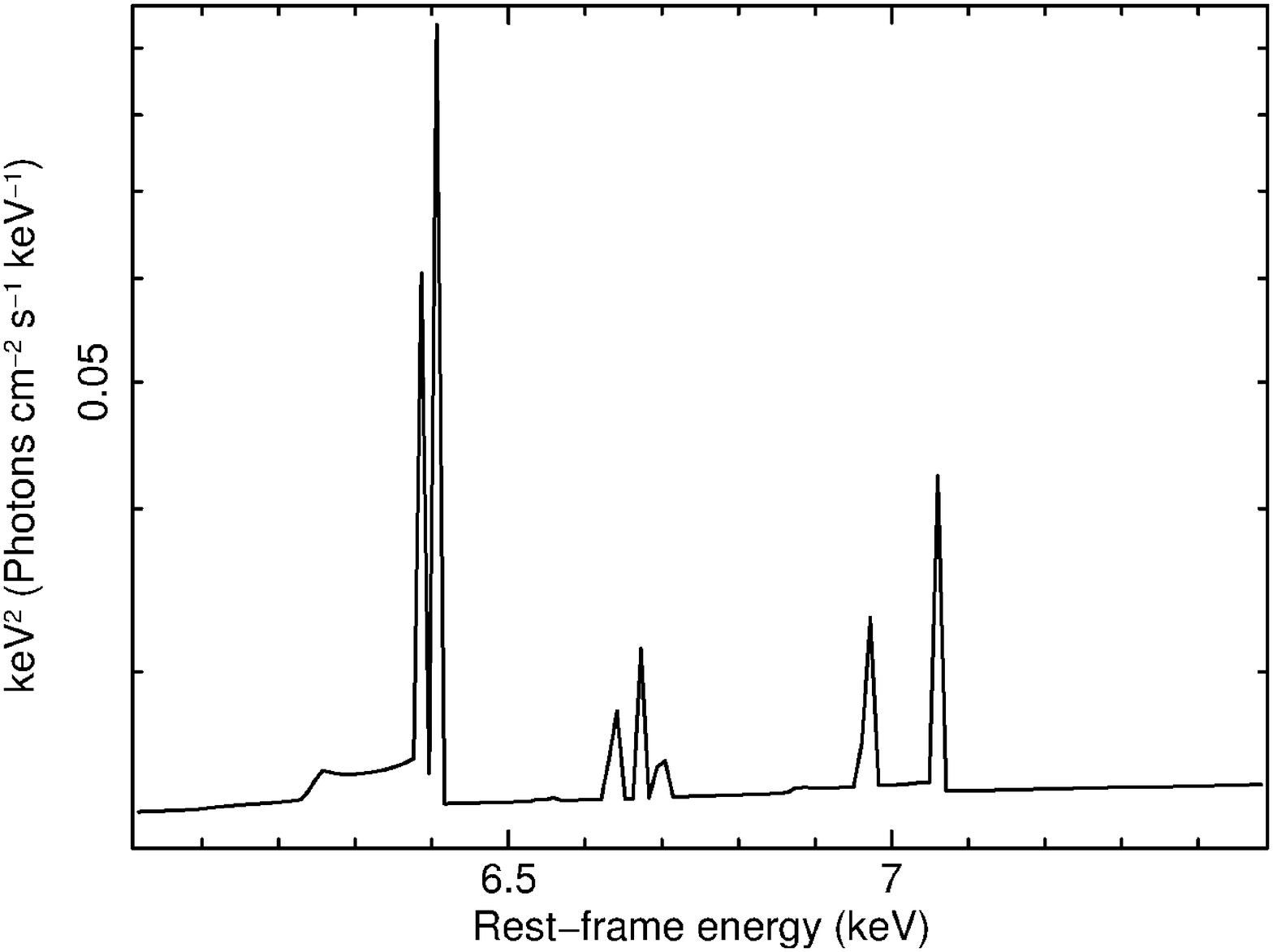}
   \caption{\emph{Upper panel:} data to model ratio of the \emph{Chandra} HEG spectrum of
     3C~120 in the E$=$5--7.5~keV energy band with respect to a
     Galactic absorbed power-law continuum model. The data are binned
     to the FWHM resolution. The vertical ines indicate the expected
     energies of the major Fe K transitions. The vertical arrow
     indicates the possible blue-shifted Fe K absorption residuals. \emph{Lower panel:}
     best-fit model including emission from cold reflection and
     photoionized gas.}
    \end{figure}

We considered also a non-equilibrium ionization collisional plasma
model, \emph{nei} in \emph{XSPEC}. This model allows to approximate the
condition in shocks or hot plasmas and is applicable also to supernova
remnants (e.g., Borkowski, Lyerly \& Reynolds 2001). This provides a higher fit improvement of $\Delta C =
13/3$ compared to a simple \emph{apec} model. We estimate a broadening
of FWHM$=$$2,400^{+1,400}_{-1,800}$~km~s$^{-1}$, a temperature of
$kT$$=$$(0.7^{+0.9}_{-0.1})$~keV, an ionization timescale of
$\tau$$=$$(1.3^{+1.2}_{-1.0})\times 10^{11}$~s~cm$^{-3}$ and a
normalization of $N$$=$$(1.2^{+0.6}_{-0.8})\times 10^{-4}$.  Equivalent results are
obtained using the plane-parallel shocked plasma model \emph{pshock}
in \emph{XSPEC}. 

%The soft X-ray emission lines indicate the presence of hot diffuse gas
%with a temperature of $kT$$\simeq$0.5~keV. This temperature is
%consistent with the typical value observed for the hot ISM in
%elliptical galaxies (e.g., Werner et al.~2009; Kim \& Pellegrini
%2012). 

The normalization of both the \emph{apec} and \emph{nei} models in
\emph{XSPEC} is defined as $N = 10^{-14}(EM/4\pi[D_A (1+z)]^2)$, where $D_A
\simeq 3.9\times 10^{26}$~cm is the angular diameter distance to
3C~120. The emission measure $EM$ is defined as $EM = \int n_e n_H
dV$, where $n_e$ and $n_H$ are the electron and hydrogen number
densities in cm$^{-3}$, respectively. Given the normalization of the
\emph{nei} component, we can estimate an emission measure of $EM
= 2\times 10^{68}N = (2.4^{+1.2}_{-1.6})\times 10^{64}$ cm$^{-3}$.

%This provides a poor fit improvement of only $\Delta
%C/\Delta\nu$$=$$2/2$ for a temperature of $kT$$\simeq$0.2~keV. Leaving
%the redshift of this component free to vary, we obtain a better fit
%required at the 99\% level ($\Delta C/\Delta\nu$$=$$12/3$) with a
%temperature of $kT$$=$$0.12\pm0.04$~keV and normalization
%$A$$>$$3$. However, this model would require a redshift of
%$3,400^{+30}_{-90}$~km~s$^{-1}$ with respect to the rest-frame of the
%source, which would indicate a possible inflow. This interpretation seems unlikely.

We also investigated the possible interpretation of the
emission lines as due to a photo-ionized component using the
\emph{xstar} model. We assume the same \emph{xstar} emission table
used for the Fe K band, considering a power-law continuum of $\Gamma =
1.7$ and turbulence velocity of 100~km~s$^{-1}$. We also considered a
normalization of $N$$\simeq$$2\times 10^{-4}$ consistent with the
source parameters assuming a covering fraction of unity.

The inclusion of this component provides a lower fit improvement compared to the
previous ones of $\Delta C/\Delta\nu$$=$$7/2$. The best-fit ionization parameter is
log$\xi$$=$$2.15^{+0.17}_{-0.24}$~erg~s$^{-1}$~cm, and the lower limit
on the column density is $N_H$$>$$1 \times 10^{20}$~cm$^{-2}$,
respectively. In this case the feature at the energy of E$\sim$777~eV
would be interpreted as O~VIII He$\beta$. However, this modeling does
not provide an adequate characterization of the emission feature at
E$\sim$719~eV because in this case the closest feature would only be O~VII RRC at E$\simeq$739~eV.   

We note that similar soft X-ray features were also reported in the
\emph{XMM-Newton} RGS spectrum of 3C~120 performed by Torresi et
al.~(2012). The authors tested two possible scenarios: Fe~XVII from
a collisionally ionized gas or O~VII RRC from a photoionized
gas. Torresi et al.~(2012) concluded that the limited signal-to-noise
and complexity of the data could not favor neither one
of these two possibilities, leaving the origin of the emitting gas as an
open question. Moreover, Tombesi et al.~(2014) reported the detection
of an excess emission in the soft X-ray band using \emph{Suzaku},
but the feature was phenomenologically parameterized with a simple blackbody component with a
temperature $kT$$\simeq$0.4~keV given the insufficient energy
resolution of that detector.

Finally, we tested an alternative interpretation of the soft X-ray
emission lines as due to L-shell fluorescence from neutral
or slightly ionized Fe. Fe L-shell fluorescence lines would be
expected at the energies of E$=$705~eV, E$=$719~eV, and E$=$792~eV,
which are roughly consistent with the observed features. 

%Considering only atomic physics, the fluorescence yield of the Fe
%L-shell is about 2\% of the Fe K-shell. Therefore, given the
%EW$\simeq$30~eV of the Fe K$\alpha$ emission line, we would expect an
%Fe L-shell emission with equivalent width of just a few eV or
%less. This would be too low to explain the observed features.
%However, if we consider that there are three times as many
%electrons that can give rise to the L-shell compared to the K-shell,
%then it may then be plausible that the EW would be higher.

Although this being an intriguing possibility, when we consider a
physical model for the reflection off cold material such as
\emph{pexmon} or \emph{MYTorus} we note that the low energy photons are
drastically affected by photoelectric absorption from the same medium
that should give rise to the Fe L lines. Therefore, neutral Fe L fluorescence
features would be undetectable in this physical scenario. 

Instead, Fe L-shell resonance emission lines may be more intense if we
consider lowly or mildly ionized reflection. In the previous section
we discussed a fit using the \emph{xillver} model which can provide a
relatively good modeling of the Fe K line with very low ionization
log$\xi$$<$0.8~erg~s~cm$^{-1}$. However, checking the region around
E$=$0.7--0.75~keV we find no detectable features from the model
because ionized Fe L-shell resonance lines, along with many others
unobserved features, would show up only for much higher ionization levels around log$\xi$$\sim$2~erg~s~cm$^{-2}$.

Therefore, an origin of the prominent emission feature at
E$\sim$0.7-0.75~keV from hot gas is preferred over fluorescence,
photoionization, and ionized reflection. The best-fit model is shown
in the lower panel of Fig.~2.

\subsubsection[]{Absorption in the soft X-ray and Fe K bands}

We performed a search for possible absorption features in both the
soft X-rays and in the Fe K band. In the soft X-ray band we do not
find absorption lines with a statistical significance higher than 99\%
indicating that, if present, the WA is either highly ionized or it has
a low column density. This is consistent with previous studies using
\emph{XMM-Newton} RGS (Ogle et al.~2005; Torresi et al.~2012). Using
an \emph{xstar} photo-ionized absorption table and assuming the typical
ionization of the WA detected in other BLRGs of log$\xi$$\simeq$2.5
erg~s$^{-1}$~cm (Reeves et al.~2009; Torresi et al.~2010, 2012) we
estimate an upper limit of the column density of just $N_H$$<$$5\times 10^{19}$~cm$^{-2}$.

\floattable 
\begin{deluxetable*}{cccccc}[t!]
%\begin{deluxetable}{lcccc}
\tablecaption{Best-fit parameters of the emission lines in 3C~120.}
\tablehead{
\colhead{E} & \colhead{$\sigma$} & \colhead{$I$} & \colhead{EW} & \colhead{ID} & \colhead{$\Delta{C}$}
}
\startdata
\multicolumn{6}{c}{soft X-ray emission lines}\\
\hline
$719\pm3$ & $8\pm2$ & $23.0\pm8.0$ & $8\pm2$ & Fe~XVII & 12\\[+4pt]
$777^{+2}_{-7}$ & $<50^a$ & $5.8\pm3.4$ & $2\pm1$ & Fe~XVIII/O~VIII & 5\\
\hline
\multicolumn{6}{c}{Fe K emission lines}\\
\hline
$6232_{-23}^{+8}$ & $<330^a$ & $0.5_{-0.2}^{+0.3}$ & $8\pm4$ & Fe~K$\alpha_c$ & 6\\
$6394\pm5$ & $<20^a$ & $1.9\pm0.3$ & $32\pm6$ & Fe~K$\alpha$ & 40\\
$6703^{+5}_{-20}$ & $<40^a$ & $0.6^{+0.4}_{-0.2}$ & $11\pm5$ & Fe~XXV He$\alpha$ & 6\\
$7055^{+21}_{-15}$ & $<80^a$ & $1.0\pm0.4$ & $20^{+6}_{-8}$ & Fe~K$\beta$ & 9\\
\enddata
\tablenotetext{a}{90\% upper limit.}
%\tablenotetext{c}{Fe K$\alpha$ Compton shoulder.}
\tablecomments{Columns: rest-frame energy in eV; line width in eV;
  intensity in units of $10^{-5}$ ph~s$^{-1}$~cm$^{-2}$; EW in eV;
  line identification; C-statistics improvement after including the Gaussian line. The line identifications are derived from the National Institute of Standards and Technology (NIST) database at www.nist.gov, Verner et al. (1996) and Kaye \& Laby at www.kayelaby.npl.co.uk.}
\end{deluxetable*}

Absorption lines indicative of ultrafast outflows in the Fe K band have been reported in two \emph{Suzaku} observations of 3C~120 performed in 2006 and 2012 (Tombesi et al.~2010b, 2014). The outflow in the 2006 \emph{Suzaku} observation was detected in the energy range between E$\simeq$7--7.3~keV with velocity $v_{out}$$\simeq$0.08c, ionization parameter log$\xi$$\simeq$3.8~erg~s$^{-1}$~cm and column density $N_H$$\simeq$$1\times 10^{22}$~cm$^{-2}$ (Tombesi et al.~2010b). The 2012 observation showed an outflow at the energy of E$\simeq$7.7~keV with higher velocity and ionization, i.e., $v_{out}$$\simeq$0.16c, log$\xi$$\simeq$4.9~erg~s$^{-1}$~cm and $N_H$$>$$2\times 10^{22}$~cm$^{-2}$ (Tombesi et al.~2014). 

The signal-to-noise and limited high-energy bandwidth of the
\emph{Chandra} HETG are not adequate to clearly detect such
features. Nonetheless, in Fig.~3 there is a hint for absorption
features in the energy range between E$\simeq$7.2--7.4~keV. We checked
for the presence of an outflow using an \emph{xstar} photo-ionization
table with an input power-law continuum slope of $\Gamma$$=$1.7 and a
turbulence velocity of 1,000 km~s$^{-1}$. We find that an outflow may
be present at the 99\% level using the F-test ($\Delta
C/\Delta\nu$$=$11/3). However, we note that the significance may be lowered down to about
90-95\% considering a blind line search between E$=$7.0-7.5~keV,
which corresponds to about 10 intervals at the HEG resolution
of 50~eV.  The best-fit indicates an outflow velocity of
$v_{out}$$=$$0.068\pm0.002$c, ionization parameter
log$\xi$$=$$3.48^{+0.17}_{-0.10}$ erg~s$^{-1}$~cm, and column density
$N_H$$=$$(3.3^{+2.4}_{-1.6})\times 10^{21}$~cm$^{-2}$. These
parameters would suggest the presence of an ultrafast outflow with
characteristics similar to the one detected in the 2006 \emph{Suzaku}
observation (Tombesi et al.~2010b, 2014). The final best-fit model is shown in Fig.~4 and the
parameters are listed in Table~3.

%*******
%THEN, WITH 3C111, CALCULATE PLOT!!! COMPARE UFO WITH WA IN RADIO GALAXIES! Jet ANGLE!!!!
%********

\section[]{Discussion}

The analysis of the \emph{Chandra} HETG spectrum of 3C~120 reveals a
rather complex multi-phase and multi-scale environment in this radio
galaxy. In the following we will describe a physical interpretation for each component.

\subsection[]{Emission in the soft X-rays}

The soft X-ray band of 3C~120 below E$\simeq$1~keV is rather complex
and the current \emph{Chandra} data show a series of emission lines in the energy
range between E$\simeq$0.6-0.9~keV. %The association of these features is ambiguous. 

The first report of possible thermal emission lines in the soft X-ray
spectrum of 3C~120 was published by Petre et al.~(1984) using \emph{Einstein} Observatory
data. The excess emission between E$=$0.5--2~keV was modeled with a
thermal bremsstrahlung component with temperature near 1~keV. The
signature of this component was the presence of weak emission lines at
energies consistent with Fe L, Mg K, and Si K. Similar soft X-ray
structures were also reported in the \emph{XMM-Newton} and \emph{Suzaku} spectra of 3C~120 (Torresi et
al.~2012; Tombesi et al.~2014). The most intense of these features are
likely associated with ionized Fe L transitions.

We tested several different alternatives for the origin of the
emission in the soft X-rays and found that a non-equilibrium
ionization collisional plasma model with an electron temperature of 
$kT$$=$$(0.7^{+0.9}_{-0.1})$~keV provides both the best-fit and a
physically plausible explanation. This model allows to approximate the condition in shocks or hot plasmas.

The fact that the data require a broadening of this hot emission component indicates
that there is an additional component compared to just the
thermal one, that for this $T \simeq 10^7$~K plasma would be only
$v_{th} = \sqrt{2kT/m_{Fe}} \simeq$50~km~s$^{-1}$. 
Instead, the width of the emission component is much larger and
corresponds to a velocity broadening of
FWHM$=$$2,400^{+1,400}_{-1,800}$~km~s$^{-1}$. This additional velocity
component could be either associated a spherical
outflow or an expanding spherical shock. We note that this velocity is
much higher than just the thermal broadening or turbulence from hot
interstellar medium, being both typically less than 100~km~s$^{-1}$ (Werner et al.~2009).

%\newpage
%\floattable 
%\begin{deluxetable}{lc}[t!]
\begin{deluxetable}{lc}
\tablecaption{Best-fit model of the \emph{Chandra} HETG spectrum of
  3C~120.}
\tablehead
{
\multicolumn{2}{c}{Power-law continuum}
}
\startdata
$\Gamma$  & $1.771\pm0.008$\\
\hline\hline
\multicolumn{2}{c}{Neutral abs}\\
\hline
log$N_H$ (cm$^{-2}$) & $20.67\pm0.05$\\
\hline\hline
\multicolumn{2}{c}{Gsmooth $\times$ nei emission}\\
\hline
FWHM ($10^3$ km~s$^{-1}$) & $2.4^{+1.4}_{-1.8}$\\
kT (keV) & $0.7^{+0.9}_{-0.1}$\\
$\tau$ ($10^{11}$ s~cm$^{-3}$) & $1.3^{+1.2}_{-1.0}$\\
EM ($10^{64}$ cm$^{-3}$) & $2.4^{+1.2}_{-1.6}$\\
\hline\hline
\multicolumn{2}{c}{MYTorus}\\
\hline
$\theta$ (deg) & 20\\
log$N_H$ (cm$^{-2}$) & $>24.8^*$\\
\hline\hline
\multicolumn{2}{c}{Xstar emission} \\
\hline
log$\xi$ (erg~s$^{-1}$~cm) & $3.75^{+0.27}_{-0.38}$\\[+4pt]
$N_H$ ($10^{21}$~cm$^{-2}$) & $>10^*$\\
\hline\hline
\multicolumn{2}{c}{Xstar absorption} \\
\hline
log$\xi$ (erg~s$^{-1}$~cm) & $3.48^{+0.17}_{-0.10}$\\[+4pt]
$N_H$ ($10^{21}$~cm$^{-2}$) & $3.3^{+2.4}_{-1.6}$\\
v$_{out}$ (c) & $0.068\pm0.002$\\
\hline\hline
\multicolumn{2}{c}{C-statistic}\\
\hline
C/$\nu$ & 2256/2209\\
\hline\hline
\multicolumn{2}{c}{Flux$^a$} \\
\hline
0.5--2~keV  & 3.4\\
2--7~keV   & 4.1\\
\enddata
\tablenotetext{a}{Intrinsic power-law continuum flux in units of $10^{-11}$ erg~s$^{-1}$~cm$^{-2}$}
\tablenotetext{*}{90\% lower or upper limit.}
%\tablecomments{ }
\end{deluxetable}

We estimate an emission measure of $EM = (2.4^{+1.2}_{-1.6})\times 10^{64}$ cm$^{-3}$.
Values of the emission measure of the order of $10^{64}$ cm$^{-3}$
have also been
reported also for the hot interstellar medium observed through Fe XVII
emission at temperatures of $kT$$\le$0.8~keV in the cores of large elliptical
galaxies (Werner et al. 2009). Considering the emission measure and assuming a fully ionized plasma,
$n_e \simeq 1.2 n_H$, 
and considering a roughly spherical volume of
radius $\sim$1.3~kpc corresponding to the extent of the region observed
with the \emph{Chandra} HETG of $\simeq$2 arcsec, we then estimate
a number density for the gas of $n_H \sim 0.5$ cm$^{-3}$. The temperature, density and extension of the gas are
indeed consistent with the values expected for the hot interstellar
medium in giant elliptical galaxies (Xu et al. 2002; Werner et
al. 2009; Kim \& Pellegrini 2012). 

Evidence for an excess
of emission in the \emph{Chandra} image from 1 arcsec to 3 arcsec with respect to
the point spread function reported in Ogle et al.~(2005) is consistent
with extended soft X-ray emission. We note that the resolving power of the grating may decrease if the
source is extended. For instance, from Fig.~8.23 of the \emph{Chandra}
proposers
guide\footnote{http://cxc.cfa.harvard.edu/proposer/POG/html/chap8.html
  \#tth\_sEc8.2}
we see that the MEG resolution could degrade down to $E/\Delta
E$$\simeq$200 for a 5~arcsec extended source at the wavelength of
17~\AA. Therefore, at the energy of E$\simeq$700~eV the resolution may
degrade down to a worst value of 3.5~eV. However, we note that the observed emission
line at E$\simeq$720~eV listed in Table~2 is resolved with a width
$\sigma$$=$$8\pm2$~eV, which is at least twice larger than the worse
energy resolution. Subtracting the 3.5~eV resolution in quadrature may
only slightly reduce the broadening to around 7~eV, which is still consistent
within the 1~sigma errors. Moreover, the same line, with consistent broadening, was reported also
in the \emph{XMM-Newton} RGS spectrum by Torresi et al.~(2012). Thus,
we are confident that the broadening of the line is not affected by
the soft X-ray emission region possibly extending up to a few
arcsecs. 

The total X-ray luminosity between E$=$0.1--10~keV estimated for the
hot gas in 3C~120 is
$L_{hot}$$\simeq$$1.5\times10^{42}$~erg~s$^{-1}$. In order to keep the
gas hot and prevent a catastrophic cooling flow, there must be some
source of energy. An order of magnitude estimate of the cooling
timescale for the hot gas associated with the galaxy can be estimated
as $t_{cool} = U/L_{hot}$, where $U$ is the internal energy of the gas
defined as $U = (3/2)NkT$, where $N$ is the number of particles and
$T$ is the temperature. For an estimated density of $n$$\simeq$0.5
cm$^{-3}$ and considering a spherical region of radius about 1.3~kpc
and temperature $T$$\simeq$$10^7$~K, we derive $U \simeq 6\times
10^{55}$ ergs. We can then estimate a short cooling time of the order
of $\sim$1~Myr or up to $\sim$10~Myr using the prescription in
Peterson \& Fabian (2006). Therefore, either most of the gas was heated more
recently than this time-scale or there is an additional source of
heating that keeps it hot. 

Shocks or outbursts can disturb the typical collisional plasma found
in the galactic interstellar medium. The spectrum emitted by this plasma contains
diagnostics that have been used to determine the time since the
disturbing event, although this determination becomes uncertain as the
elements return to equilibrium. For instance, this method has been
used to determine the ’age’ of supernova remnants or to determine 
when individual knots in the ejecta were shocked (e.g., Hwang \& Laming 2003; Yatsu et al.~2005). 

From our best-fit with a non-equilibrium ionization model we estimate
an ionization timescale of $\tau$$=$$(1.3^{+1.2}_{-1.0})\times 10^{11}$~s~cm$^{-3}$.
Figure~1 in Smith \& Hughes (2010) show the typical timescales of
different ions to reach equilibrium for different temperatures. For a
temperature of $T$$\simeq$$10^7$~K estimated here from the Fe L
emission lines we see that a characteristic time-scale to reach 90\%
of equilibrium would be much longer, of $\tau$$\simeq$$9\times10^{11}$~s~cm$^{-3}$. Therefore, the observed
gas is about a factor of seven in time away from reaching equilibrium. 

 Considering the indicative number density of $n$$\simeq$0.5~cm$^{-3}$, we
 obtain a timescale of $\sim$$10^{11}$~s or about 10,000 years since the main
 disturbance of the gas occurred. It is tempting to relate such
 time-scale with the onset or with intermittent outburst activity of
 the powerful jet or winds in this AGN, perhaps in line with evolutionary scenarios
 (Reynolds \& Begelman 1997; Czerny et al.~2009; Baldi et al.~2015). 
This scenario is also consistent with the possible thermal emission
from isothermal shocks suggested by Halpern et al.~(1985) as
an explanation of the continuum emission and variability of 3C~120
using data from the \emph{Einstein} X-ray observatory. The shocks in
Halpern et al.~(1985) have higher density, velocity ($v_s \sim
20,000$~km~s$^{-1}$), and temperature ($kT \sim 7\times 10^7$~K), and
are located closer to the black hole (at $\sim$pc scales) compared to
the ones observed through Fe L line emission here and in Petre et
al.~(1984). In analogy to supernova
 remnants, this may indicate a superposition or stratification of AGN shocks with
 various locations and ages.

  \begin{figure}
  \centering
   \includegraphics[width=8.5cm,height=7cm,angle=0]{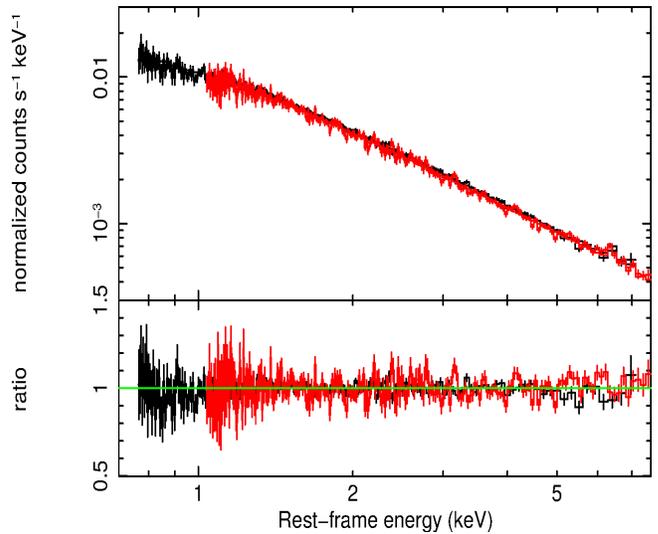}
   \caption{Combined \emph{Chandra} MEG (black) and HEG (red) spectra
     of 3C~120 compared with the best-fit model. \emph{Upper panel:} spectra and continuum model, the data
     are divided by the response effective area for each channel. \emph{Lower panel:} data to model
     ratio. The data are binned to 4$\times$ the FWHM resolution and
     to a minimum signal-to-noise of 5 for clarity.}
    \end{figure}

Further support for the hot gas being heated by AGN driven shocks comes also from
the comparison of the warm mid-IR H$_2$ emission compared
to the thermal emission in radio galaxies. In particular, Ogle et
al.~(2010) detected the H$_2$ S(1) line in 3C~120, suggesting that the
most likely heating mechanism is jet-driven shocks in a multiphase
interstellar medium. Moreover, Lanz et al.~(2015) reported a similar luminosity in the mid­IR
H$_2$ lines and in the thermal X­ray emission for radio galaxies whose mid­IR
spectrum indicates the presence of shocks.

These evidences, together with the large velocity broadening of the hot
emission component in 3C~120 of about 1,000 km~s$^{-1}$ point to
the possibility that the velocity width could be due to a spherical
outflow or an expanding spherical shocked bubble, perhaps driven by
the winds or jet from the central AGN.
We note that the Fermi bubbles observed in our Milky Way have indeed a
comparable temperature and size, and they are estimated to be
expanding with a shock speed of $v_{shock}$$\simeq$1,000~km~s$^{-1}$ (Fox et al.~2015;
Nicastro et al.~2016).

We note that 3C~120 shows both ultrafast outflows and radio
jets with powers of $P_{UFO} \sim 1\times 10^{44}$~erg~s$^{-1}$ and
$P_{jet} \sim 5\times 10^{44}$~erg~s$^{-1}$, respectively (Tombesi et
al.~2010b, 2014; Torresi et al.~2012). Hot shocked bubble are indeed expected from theory of AGN
feedback driven by winds or jets (e.g., Wagner et al.~2012, 2013;
Zubovas \& Nayakshin 2014).
As the AGN wind sweeps up the ambient medium the forward shock
decelerates producing cooler gas with a temperature of $T \simeq
10^7$~K for a shock velocity of $v_{shock}$$\simeq$1,000~km~s$^{-1}$. 
Momentum conservation implies that the mean density swept up by the
AGN wind is then of the order of $n$$\simeq$0.1~cm$^{-3}$ for an AGN luminosity of $L
\sim 10^{45}$~erg~s$^{-1}$, a shock radius of $R_s\sim 1$ kpc and
velocity of $v_{shock}$$\sim$1,000~km~s$^{-1}$. The typical X-ray luminosity
of such wind shocked gas from bremsstrahlung would be of $L_{X} \sim
10^{42}$~erg~s$^{-1}$ (Bourne \& Nayakshin 2013; Nims et al.~2015). These parameters are indeed consistent with the estimates for the hot X-ray emitting gas in 3C~120.

A possible variability of the emitter was reported by Petre et
al.~(1984). Torresi et al.~(2012) reported the following parameters
for the emission line in the \emph{XMM-Newton} RGS spectrum of 3C~120
obtained in 2003: line energy of E$\sim$730~eV, flux of $I \simeq
7\times 10^{-5}$~ph~s$^{-1}$~cm$^{-2}$, and line width
FWHM$=$$4,800^{+2,300}_{-2,700}$~km~s$^{-1}$. Comparing to the
parameters of the emission line in our \emph{Chandra} HETG spectrum
reported in Table~2, we note that the main difference in the 11 years
spanning between the two observations is a factor of three increase in
line flux. In the framework of a hot shocked bubble, we would expect
an increase in X-ray bremsstrahlung luminosity as the wind shock
expands to larger radii (e.g., Nims et al.~2015). Future high-energy
resolution observations of the source may confirm this scenario. 

On the other hand, we note that the X-ray emission expected from star formation can be
estimated as $L_{x(0.5-8keV)}$$\simeq$$4\times 10^{39} (\dot{M}_*/1 M_{\odot}
yr^{-1})$~erg~s$^{-1}$, with a scatter of $\simeq$0.4~dex, where $\dot{M}_*$ is the star formation rate
(Mineo et al.~2014). Given the star formation rate in 3C~120 of
  $\dot{M}_* = 2.8^{+0.9}_{-1.0}$~$M_{\odot}$~yr$^{-1}$ estimated by
  Westhues et al.~(2016) using multiwavelegth data, the resultant X-ray luminosity would be orders of magnitude lower than observed.

The radio galaxy 3C~120 is an isolated, peculiar S0 galaxy about
25~kpc in diameter. The optical appearance is complex, with multiple dust lanes
and two suggestive spiral arms that become radial features at a
projected distance of about 6.3~kpc from the core. It is not clear
whether the peculiar optical morphology, perturbed rotation curve and wide-spread enhanced star formation are due to the
effects of tidal perturbation or interaction of the radio jet with the
galaxy (Harris et al.~2004).  It has been suggested that the optical data indicate a merger
which is near completion (Heckman et al. 1986; Moles et al. 1988).

Indeed, integral field spectroscopy of the central regions of 3C~120 shows
evidence of shells in the central kpc region that may be remnants of a
past merging event in this galaxy and/or interaction of the radio get
with the intergalactic medium (Garcia-Lorenzo et al.~2005). 3C 120 is a bulge-dominated galaxy that has, most probably,
experienced a merging event with a less massive galaxy. That galaxy was completely disrupted in the merging process, falling
in parts that produce, most probably, many of the observed optical
structures. There is also evidence that a substantial fraction of its gas has been channeled toward the
inner regions, perhaps activating a new feeding phase for the central
supermassive black hole (Garcia-Lorenzo et al.~2005).

Indeed, CO emission data add substantial support to the hypothesis that 3C~120 involves
a merger of gas-rich galaxies. The CO line profile is broad with
FWHM$=$500~km~s$^{-1}$ and full width at zero intensity (FWZI) of 710~km~s$^{-1}$ (Mazzarella et al.~1993;
Evans et al.~2005). The velocity broadening is probably dominated by
rotation, but there is also a component coming from disturbances due
to the radio jet or AGN wind. The estimated H$_2$ mass is large, log$M(H_{2})$$=$9.79~$M_{\odot}$. This corresponds to about twice the
molecular gas content of the Milky Way. These CO data corroborate evidence from optical imaging that this and
similar radio galaxies originate in collisions and mergers between
disk galaxies, as recently suggested also by other authors (e.g.,
Chiaberge et al.~2015).  

Combined radio,
optical and \emph{Chandra} X-ray images show complex interactions and
shocks between the radio jet and colder material. Harris et al.~(2004)
showed that the radio jet may be impacting a cold, spiral-like structure
observable at 4 arcsec (about 2.5~kpc) from the nucleus. A bright
radio, X-ray and optical jet knot emission is observed at this
location (Harris et al.~2004). Evidence for complex mixing between hot shocked gas
and cold gas due to the interaction between AGN winds or jets has been
recently reported in an increasing number of sources (e.g., Morganti
et al.~2013; Tadhunter et al.~2014; Tombesi et al.~2015; Feruglio et al.~2015;
Dasyra et al.~2016). All these characteristics may place 3C~120 in a comparable parameter space of the AGN dominated ULIRGs (Veilleux et al.~2013; Cicone et al.~2014).

\subsection[]{Lack of a warm absorber}

We do not detect any significant ionized absorption feature in the soft X-rays. This is consistent
with previous studies using \emph{XMM-Newton} RGS (Ogle et al. 2005;
Torresi et al. 2012) and a shorter \emph{Chandra} HETG (McKernan et al.~2007). 
Using an \emph{xstar} photo-ionized absorption table and assuming the typical ionization of the WA detected in
other BLRGs of log$\xi$$\simeq$2.5~erg~s$^{-1}$~cm and a velocity consistent with
zero at the source rest-frame (Reeves et al. 2009; Torresi et
al. 2010, 2012), we estimate a very low upper limit of the column
density of a possible WA of just $N_H$$<$$5\times10^{19}$~cm$^{-2}$.

From the previous discussion of hot emission in the soft X-rays, we
suggest that the lack of a warm absorber may be due to the fact that
the temperature of the hot interstellar medium of $T\sim 10^7$~K is
much higher than the typical one for warm absorbers of $T \sim 10^5$~K
(Chakravorty et al.~2009). Therefore, it is plausible that the warm
absorbing gas often observed in Seyfert galaxies, if present, has been heated up too
much and it is not observable.

\subsection[]{Fe K emission and absorption}

The Fe K band spectrum of 3C~120 in Fig.~2 is very rich, showing a series of at
least four narrow emission and absorption lines due to iron at different ionization
states. An unresolved Fe K$\alpha$ emission line at E$\simeq$6.39~keV,
an associated Compton shoulder at E$\simeq$6.23~keV and the Fe
K$\beta$ at E$\simeq$7.05~keV from cold or neutral gas. 

The Fe K$\alpha$ emission line is unresolved, with a FWHM of less than
2,300~km~s$^{-1}$, consistent with previous estimates (Yaqoob et al.~2004; Shu et al.~2010). 
This value is comparable to the width of the core of the broad
line region (BLR) observed in optical H$\alpha$ and H$\beta$ lines
(Kollatschny et al.~2014), possibly suggesting a spatial overlap
between the cold X-ray material and optical line emitting gas. The
narrowness of the line may be also consistent with the low inclination
observed from the radio jet of 20$^{\circ}$ (Jorstad et al.~2005). 

It is interesting to note that although the FWHM of the
H$\alpha$/H$\beta$ lines is about 2,000~km~s$^{-1}$, the FWZI calculated considering the width of the line at
which it reaches the continuum level is much larger, reaching
velocities of up to 15,000~km~s$^{-1}$ (Phillips \& Osterbrock 1975; Osterbrock 1977).
The large difference between the FWHM and the FWZI indicates that the
lines should have broad and complex wings. The FWZI is indeed more
representative of the full range of velocities in the BLR. Such broad
wings could possibly be associated with the part of the BLR closer to
the black hole if due to rotation and/or to a possible fast bipolar
outflow component.
If the Fe K$\alpha$ at E$\simeq$6.4~keV is related to the BLR, then it may show
similar broad wings. However, the signal-to-noise of the current
\emph{Chandra} HETG observation is not high enough to test this possibility. 

A detailed characterization using models of X-ray reflection from cold
material gives a low reflection fraction of $R$$=$$0.22\pm0.04$
consistent with the relatively low EW of the Fe K$\alpha$ emission
line of EW$=$$32\pm6$~eV. The column density of the torus is estimated
to be Compton-thick, with $N_H$$>$$6\times 10^{24}$~cm$^{-2}$ in the
equatorial plane if the \emph{MYTorus} geometry is assumed. The
detection of a Compton shoulder to the main Fe K$\alpha$ line is
consistent with these estimates (Awaki et al.~2008). The overall low reflection fraction
may indicate a clumpy torus outside the line of sight or a high
opening angle of the reflecting material (e.g., Tazaki et al.~2013).   

Comparing the extrapolated E$=$2-10~keV X-ray luminosity of $L_{2-10}
\simeq 1.4\times 10^{44}$~erg~s$^{-1}$ and EW$=$$32\pm6$~eV derived
here for 3C~120 with the Fig.~7 in Ricci et al.~(2013) we see that
the values for this radio galaxy are consistent with those expected in the range between Seyferts
and quasars. This is consistent with the X-ray Baldwin effect, which
would relate the decrease of the covering factor of the torus with the
increase in luminosity (e.g., Iwasawa \& Taniguchi 1993). 

The shape and strength of the neutral Fe K$\alpha$ emission line
suggest that the material feeding the accretion disk, or the torus, may be
in the form of Compton-thick, clumpy clouds in an equatorial
distribution. This is consistent with a recent model linking the micro
and macro properties of AGN feeding and feedback (Gaspari \& Sadowski
2017). In this model the inflow occurs via chaotic cold accretion, in which a rain of cold clouds condensing out of the quenched cooling
flow is recurrently funneled via inelastic collisions. At the
accretion disk scales, the accretion energy is transformed into
ultrafast outflows and jets, ejecting most of the inflowing mass. The
relatively high Eddington ratio of 3C~120 of $L/L_{Edd}\sim 0.3$ is
also consistent with the accretion disk being mostly relatively cold
and efficient, with episodic thermal or magnetic instabilities linked
to jet ejection events (Chatterjee et al.~2009; Tombesi et al.~2010;
Lohfink et al.~2013).

Besides emission and reflection from cold material, an Fe~XXV
He$\alpha$ emission line at E$\simeq$6.7~keV is also indicative
of the presence of a significant amount of highly ionized gas in the
central regions of this source. From the photoionization modeling we
derive an ionization parameter of log$\xi$$=$$3.75^{+0.27}_{-0.38}$~erg~s$^{-1}$~cm and a column density
of $N_H$$>1\times10^{22}$~cm$^{-2}$. From the definition of the
  ionization parameter, and assuming a compact spherical shell distribution ($\Delta R/R \le 1$), we can estimate an
 upper limit on the distance of the emitting material from the X-ray
 source as $R_{max} = L_{ion}/\xi N_H$ (e.g., Tombesi et al.~2013a;
 Gofford et al.~2015).  Substituting the estimated column density, ionization parameter, and ionizing
luminosity, we derive the upper limit of $\simeq$2~pc.

Absorption residuals in the energy range
E$\simeq$7.2--7.4~keV are suggestive of a possible ultrafast outflow
with mildly relativistic velocity of  $v_{out}$$=$$0.068\pm0.002$c, a
high ionization parameter log$\xi$$=$$3.48^{+0.17}_{-0.10}$~erg~s$^{-1}$~cm, and a column density
of $N_H$$=$$(3.3^{+2.4}_{-1.6})\times10^{21}$~cm$^{-2}$. These
parameters are consistent with the ultrafast outflow detected in a
previous 2006 \emph{Suzaku} observation (Tombesi et al.~2010b,
2014). Considering the previous formula for $R_{max}$ used for the ionized emitter, and
substituting the parameters estimated from the fit, we can derive an
upper limit of the distance of the absorber from the X-ray source of
less than $\simeq$10~pc. 

Interestingly, the ionization parameter, column density, and distance of the
Fe~XXV emitter and the ultrafast outflow are comparable, suggesting the possibility
that the ionized emission and absorption may originate from the same
or related material. Therefore, we may be observing emission and
absorption from the same extended outflow, as evinced also from
\emph{Suzaku} observations of another radio galaxy, namely 3C~111 (Tombesi et
al.~2013b). 

An intriguing phenomenon detected in a number of radio imaging
  observations of AGN jets in blazars and some radio galaxies, notably
  3C~120, is the presence of transverse gradients in the Faraday rotation across the
jet at parsec scales (G{\'o}mez et al.~2000, 2008, 2011). The origin and nature of such Faraday screen is still debated. It may be linked to the jet itself,
such as a jet sheath with an helical/toroidal magnetic field (e.g., Gabuzda et al.~2014), or to magnetized gaseous clouds
completely external to the jet (e.g., Zavala \& Taylor 2004). In
the case of 3C~120, the Faraday screen has been linked to a clumpy
sheath of thermal electrons surrounding the radio jet and interacting
with the jet at the de-projected distance of 8~pc, with a hydrogen
equivalent column density of $N_H \sim 6\times 10^{22}$~cm$^{-2}$
(G{\'o}mez et al.~2000, 2008). The fact that the observed changes in the polarization
properties of the underlying jet emission and in the rotation measure
seem uncorrelated implies that the Faraday screen and the jet are
likely not closely physically related. At the same time, sign
reversals in the rotation measure along the jet observed on the time
scale of a few/several years imply that the Faraday screen medium is
at least mildly relativistic (G{\'o}mez et al.~2011).

It is interesting to consider that the ionized emission/absorption components observed in the Fe K band
may be linked to the Faraday screen seen in radio. In fact, the column
density, distance and relatively large outflow velocity of
$v_{out}$$\simeq$0.07~c are consistent with this possibility and they
could naturally explain the observed rotation measure variability
(Zavala \& Taylor 2004).

%{\bf EXTRACT HUBBLE IMAGE 3C120!!!}

%{\bf ZEROTH ORDER CHANDA... ASK RICHARD MUSHOTZKY!!! CHRIS? JAMES?} 

%{\bf**** CHECK ZERO-TH ORDER IMAGE!!! SEE IF EXTENDED EMISSION!? SHOULD NOT BE PROBLEM PILE-UP!!! Less area than only ACIS****}

%{\bf TARO SHIMITZU!! EXTRACT GALAXY FROM IMAGE HUBBLE OF AGN GALAXY!!!}

%{\bf ****PICTURE, HOT INTERSTELLAR MEDIUM ELLIPTICAL GALAXY + LUMINOUS AGN IN THE CENTER + POWERFUL RADIO JET!!!*** UFO, WARM ABSORBER along ioniztion cone}

%{\bf ***SAY SOMETHING ABOUT HOT ISM ELLIPTICAL GALAXIES*** What is connection with AGN feedback from jets and winds? }

%{\bf COMPARE WITH PREVIOUS OBSERVATIONS!!?? VARIABILITY?}

%{\bf PLOT COMPARISON WINDS AND INCLINATION!!! compare also with UFO?}

\section{Conclusions}

We reported the spectral analysis of a long 200~ks \emph{Chandra} HETG
observation of the radio galaxy 3C~120. The spectrum shows complex
emission and absorption features in the soft X-rays and the Fe K
band. We detect a neutral Fe
K$\alpha$ line and accompanying emission lines indicative of a
Compton-thick cold reflector with low reflection fraction, possibly
due to a large opening angle of the torus. We observe also a
highly ionized Fe~XXV emission feature and possible evidence for a
highly ionized disk wind consistent with previous claims. 
We do not find evidence for a warm absorber, which may instead be replaced by a hot emitting gas with temperature $kT
\simeq 0.7$~keV observed as soft X-ray emission from ionized Fe
L-shell lines which may originate from a kpc scale shocked bubble
inflated by the AGN wind or jet with a shock velocity of about
1,000~km~s$^{-1}$. This dataset shows the wealth of information that can be derived
from high energy resolution X-ray spectroscopy and it will be very
interesting to compare to other multiwavelength observations to
explore the multi-phase and multi-scale structure of AGN feeding and
feedback of powerful radio galaxies. In particular, deeper X-ray
observations with high energy resolution spectrometers will be
fundamental to constrain the accretion onto and the ejection from the central
supermassive black hole and deeper mm or IR observations
will allow to investigate the connection with the large reservoir
molecular gas.

\acknowledgments

F.T. thanks M. Elvis, R. Petre, and S. Veilleux for the useful
discussions. F.T. and C.S.R. acknowledges
support for this work by the National Aeronautics and Space
Administration (NASA) through Chandra Award Number GO4-15103A issued
by the Chandra X-ray Observatory Center, which is operated by the
Smithsonian Astrophysical Observatory for and on behalf of NASA under
contract NAS8-03060. C.S.R. acknowledges support from the NASA ASTRO-H
grant NNX15AU54G. {\L}.\,S. was supported by the Polish National Science Centre through the grant DEC­2012/04/A/ST9/00083.

\end{document}